\documentclass[aps,pra,amsmath,reprint,superscriptaddress]{revtex4-1}
\usepackage{amsmath,txfonts,bm,color,dcolumn,graphicx}
\begin{document}

\title{An efficient solver for large structured eigenvalue problems in relativistic quantum chemistry} 

\author{Toru Shiozaki}
\email{shiozaki@northwestern.edu}
\affiliation{Department of Chemistry, Northwestern University, 2145 Sheridan Rd., Evanston, Illinois 60208, USA}
\date{\today}

\begin{abstract}
We report an efficient program for computing the eigenvalues and symmetry-adapted eigenvectors of very large quaternionic (or Hermitian skew-Hamiltonian) matrices,
using which structure-preserving diagonalization of matrices of dimension $N>10000$ is now routine on a single computer node.
Such matrices appear frequently in relativistic quantum chemistry owing to the time-reversal symmetry.
The implementation is based on a blocked version of the Paige--Van Loan algorithm [D. Kressner, BIT \textbf{43}, 775 (2003)],
which allows us to use the Level 3 BLAS subroutines for most of the computations.
Taking advantage of the symmetry, the program is faster by up to a factor of two than state-of-the-art implementations of complex Hermitian diagonalization;
diagonalizing a $12800\times 12800$ matrix took 42.8 (9.5) and 85.6 (12.6) minutes with 1 CPU core (16 CPU cores) using
our symmetry-adapted solver and Intel MKL's {\tt ZHEEV} that is not structure-preserving, respectively.
The source code is publicly available under the FreeBSD license.
\end{abstract}

\maketitle
\section{Introduction}
In relativistic quantum mechanics, the Hamiltonian operator commutes with the time-reversal operator $\hat{\mathcal{K}}$ in the absence of external magnetic fields, i.e.,
\begin{align}
[\hat{H}, \hat{\mathcal{K}}] = 0 \quad \mathrm{or}\quad \hat{\mathcal{K}} \hat{H} \hat{\mathcal{K}}^{-1} = \hat{H},
\end{align}
in which $\hat{\mathcal{K}}$ is a product of a unitary operator and the complex conjugation operator \cite{Wignerbook,SaueThesis,Reiherbook}.
In Kramers-restricted relativistic electronic structure calculations based on the four-component Dirac equation or its two-component approximations \cite{Reiherbook},
many of the matrices, including the Fock matrix and reduced density matrices, 
have the following structure due to the time-reversal symmetry and Hermicity:
\begin{align}
A = \left(
\begin{array}{cc}
D & -E^\ast \\
E & D^\ast 
\end{array}
\right), \label{matsym}
\end{align}
where $A \in \mathbb{C}^{2n\times 2n}$ and $D, E \in \mathbb{C}^{n\times n}$ with $D^T = D^\ast$ and $E^T = -E$.
In the case of the Fock matrix in two- and four-component calculations,
for instance, the values of $n$ are $N_\mathrm{bas}$ and  $2N_\mathrm{bas}$, respectively, where $N_\mathrm{bas}$ is the number of the scalar basis functions. 
This structure is referred to as Hermitian plus skew-Hamiltonian in the applied mathematics literature \cite{KressnerBook}.
Hereafter, we call this symmetry ``quaternionic" in this article, because the matrix in Eq.~\eqref{matsym}
can also be represented by an $n\times n$ matrix of quaternions.
Note that the diagonal of $E$ is zero owing to the skew symmetry, which plays an important role later.
In many of the relativistic quantum chemical algorithms, such as mean-field methods, one must efficiently diagonalize very large matrices of this kind.

Physically, this structure arises because the time reversal operator $\hat{\mathcal{K}}$ transforms molecular spinors as
\begin{subequations}
\begin{align}
& \hat{\mathcal{K}} \phi_i  = \bar{\phi}_i,\\ 
& \hat{\mathcal{K}} \bar{\phi}_i = -\phi_i,
\end{align}
\end{subequations}
where $\phi_i$ and $\bar{\phi}_i$ are said to be Kramers partners;
using them, it follows naturally for a time-reversal symmetric operator $\hat{f}$ (such as the Fock operator) that
\begin{subequations}
\begin{align}
\langle \bar{\phi}_i | \hat{f} |\bar{\phi}_j\rangle^\ast
&=\hat{\mathcal{K}} \langle \bar{\phi}_i | \hat{f} |\bar{\phi}_j\rangle
= \langle \hat{\mathcal{K}} \bar{\phi}_i |  \hat{\mathcal{K}} \hat{f} \hat{\mathcal{K}}^{-1} |\hat{\mathcal{K}}\bar{\phi}_j\rangle, \nonumber\\
&=  \langle \phi_i | \hat{f} |\phi_j\rangle \\
\langle \bar{\phi}_i | \hat{f} |\phi_j\rangle^\ast
&=\hat{\mathcal{K}} \langle \bar{\phi}_i | \hat{f} |\phi_j\rangle
= \langle \hat{\mathcal{K}} \bar{\phi}_i |  \hat{\mathcal{K}} \hat{f} \hat{\mathcal{K}}^{-1} |\hat{\mathcal{K}}\phi_j\rangle \nonumber\\
&= -\langle \phi_i | \hat{f} |\bar{\phi}_j\rangle,
\end{align}
\end{subequations}
hence the structure in Eq.~\eqref{matsym}.
The readers are referred to Ref.~\onlinecite{SaueThesis} for thorough discussions on time-reversal symmetry in quantum chemistry. 

It is trivial to show that when $(u^T v^T)^T$, with $u$ and $v$ being column vectors of length $n$, is an eigenvector of Eq.~\eqref{matsym}, 
$(-v^{\ast T} u^{\ast T})^T$ is also an eigenvector with the same eigenvalue. 
Therefore, the eigenvalue problem for $A$ can be written as 
\begin{align}
\left( \begin{array}{cc} D & -E^\ast \\ E & D^\ast \end{array} \right)
\left( \begin{array}{cc} U & -V^\ast \\ V & U^\ast \end{array} \right)
= \left( \begin{array}{cc} U & -V^\ast \\ V & U^\ast \end{array} \right)
\left( \begin{array}{cc} \epsilon & 0 \\ 0 & \epsilon \end{array} \right),
\label{eigen}
\end{align}
in which $U, V \in \mathbb{C}^{n\times n}$ and $\epsilon$ is a diagonal matrix whose elements are real.
However, there is a freedom to rotate among the eigenvectors that correspond to the same eigenvalue and multiply an arbitrary phase factor to each eigenvector,
which, in general, result in a set of eigenvectors without the structure in Eq.~\eqref{eigen}.
The standard Hermitian diagonalization solvers (e.g., {\tt ZHEEV}) ignore this quaternionic structure.

In quantum chemical algorithms, however, it is often essential to obtain the eigenvectors of the form in Eq.~\eqref{eigen}.
For instance, they are used in complete-active-space self consistent field (CASSCF) algorithms \cite{Jensen1996JCP,Bates2015JCP} to fix the optimization frame
and remove the orbital rotations among degenerate molecular spinors.
Although it is formally possible to symmetrize the eigenvectors after diagonalization,
symmetry-adaptation procedures are numerically unstable, especially for large systems, because of the double precision arithmetic used in chemical computations.
Given the recent developments of large-scale two- and four-component relativistic quantum chemistry \cite{Belpassi2011PCCP,Seino2012JCP,Peng2013JCP,Kelley2013JCP,Hrda2014JCompC,Bates2015JCP,Repisky2015JCTC,Shiozaki2015JCTC},
it is now of practical importance to develop a highly efficient tool for diagonalizing quaternionic matrices that preserves the structure.

The conventional approach to obtaining the symmetry-adapted eigenvectors of a quaternionic matrix
is to use the quaternion algebra.
By representing the matrix as an $n\times n$ matrix of quaternions \cite{Saue1999JCP},
\begin{align}
A = A^0 + A^i \breve{i} + A^j \breve{j} + A^k \breve{k}
\end{align}
with $A^i\in \mathbb{R}^{n\times n}$,
a quaternionic version of the Householder transformation or Jacobi rotation can be performed;
the algorithm has been reported several times in the literature (e.g., Ref.~\cite{Rosch1983CP,SaueThesis,DeLeo2002JMP,Sangwine2006AMC} just to name a few).
The quaternion algebra is nevertheless somewhat complicated,
and its computation cannot be easily mapped to highly optimized linear algebra libraries such as BLAS and LAPACK.

In this article, we take an alternative route
and report a highly efficient implementation of the so-called Paige--Van Loan algorithm.
The Paige--Van Loan algorithm is a generalization of the standard tridiagonalization algorithm to
skew-Hamiltonian matrices, a special case of which is a quaternionic matrix.
The blocking scheme proposed by Kressner \cite{Kressner2003BIT,KressnerBook} is used to map most of the computations to the Level 3 BLAS subroutines 
and to achieve high efficiency. 
In what follows, we first review the underlying algorithms, followed by our efficient implementation. 

\section{Algorithm}
\subsection{Paige--Van Loan algorithm}
The Householder transformation is often used in tridiagonalization of a Hermitian matrix. It transforms a Hermitian matrix $A$ as
\begin{align}
&A_1 \leftarrow H_1 A H_1^\dagger,
\label{hhtrans}\\
&H_1 = 1 - \beta_1 v_1v_1^\dagger,
\end{align}
such that the first column and row of $A_1$ is zero except for the first two elements.
The transformation matrix can be computed using the first column of the input matrix $A$ as
\begin{align}
&v_1 = \left[0,\, 1,\, (A)_{3,1}/\alpha,\, (A)_{4,1}/\alpha,\, \cdots\right]^T,
\label{house}\\
&\beta_1 = \alpha^\ast / \gamma, \quad \alpha = (A)_{2,1}+\gamma,\\
&\gamma^2 = \sum_{i=2}^{n} |(A)_{i,1}|^2,
\end{align}
in which $(A)_{i,j}$ denotes a matrix element of $A$.
The transformation matrix $H_1$ satisfies $H_1 H_1^\dagger = H_1^\dagger H_1 = 1$ (the phase of $\gamma$ is arbitrary).
Hereafter the subscript denotes the iteration number in the algorithm.
By performing this procedure recursively, one obtains a tridiagonal matrix, i.e.,
\begin{align}
\includegraphics[keepaspectratio, width=0.45\textwidth]{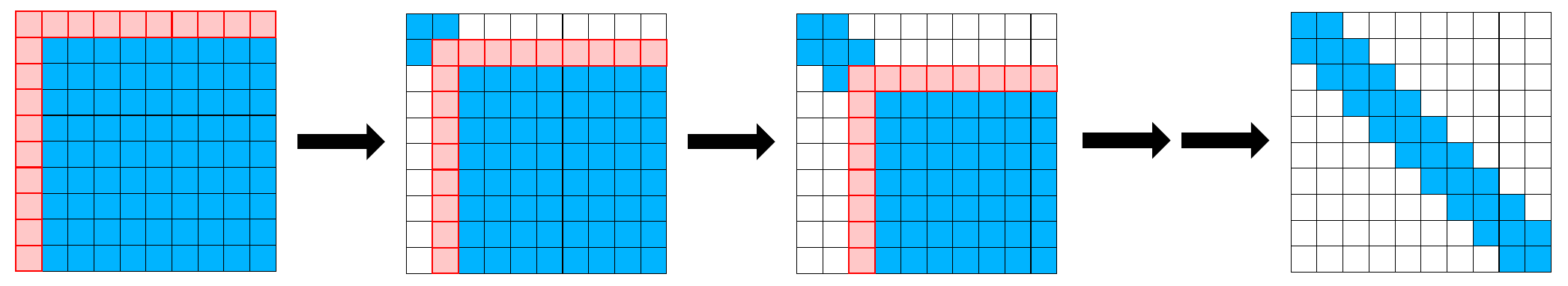}
\end{align}
When a matrix has skew symmetry $A^T = -A$, a similar procedure can be used:
\begin{align}
A_1 \leftarrow H_1 A H_1^T.
\label{hhtrans2}
\end{align}
We note in passing that the transformed matrix has the same symmetry as the original matrix.

The Paige--Van Loan algorithm \cite{Paige1981LAA,VanLoan1984LAA,KressnerBook} applies similar transformations to skew-Hamiltonian matrices
(see Ref.~\cite{Dongarra1984LAA} for its early application to quaternionic matrices).
First, a Householder transformation is applied to the off-diagonal blocks as  
\begin{align}
\includegraphics[keepaspectratio, width=0.40\textwidth]{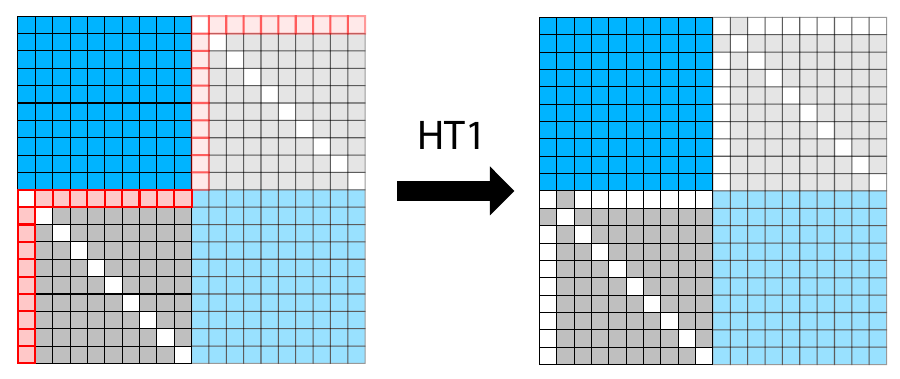}
\end{align}
Since the symmetry is preserved, the right half of the matrix does not have to be stored and transformed. 
The transformation can be written as
\begin{align}
\left( \begin{array}{cc} D & -E^\ast \\ E & D^\ast \end{array} \right) \leftarrow
\left( \begin{array}{cc} H_{E1}^\ast & 0 \\ 0 & H_{E1} \end{array}\right)\left( \begin{array}{cc} D & -E^\ast \\ E & D^\ast \end{array} \right) 
\left( \begin{array}{cc} H_{E1}^T & 0 \\ 0 & H_{E1}^\dagger \end{array}\right).
\label{ht1}
\end{align}
We then use a Givens rotation to clear out the remaining elements in the first column and row in $E$:
\begin{align}
\includegraphics[keepaspectratio, width=0.40\textwidth]{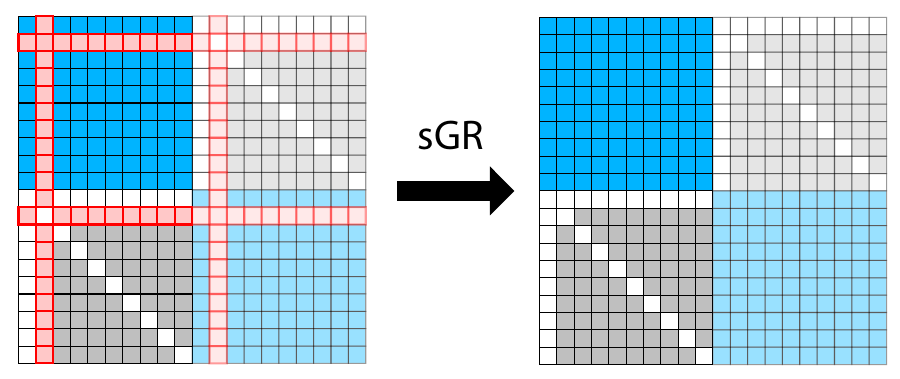}
\end{align}
The rotation is generated by
\begin{subequations}
\label{givens}
\begin{align}
&A\leftarrow G_1 A G_1^\dagger,\\
&(G_1)_{2,2} = (G_1)_{n+2,n+2} = \frac{|(A)_{2,1}|}{r} \equiv \cos\theta_1, \label{givens1}\\
&(G_1)_{2,n+2} = -(G_1)^\ast_{n+2,2} =  \frac{(A)_{2,1}(A)^\ast_{n+2,1}}{r|(A)_{2,1}|} \equiv \sin \theta_1, \label{givens2}
\end{align}
\end{subequations}
where $r = [|(A)_{2,1}|^2 + |(A)_{n+2,1}|^2]^{1/2}$,
while the other elements of $G_1$ are identical to the unit matrix.
Finally we perform another Householder transformation for the $D$ matrix, yielding 
\begin{align}
\includegraphics[keepaspectratio, width=0.40\textwidth]{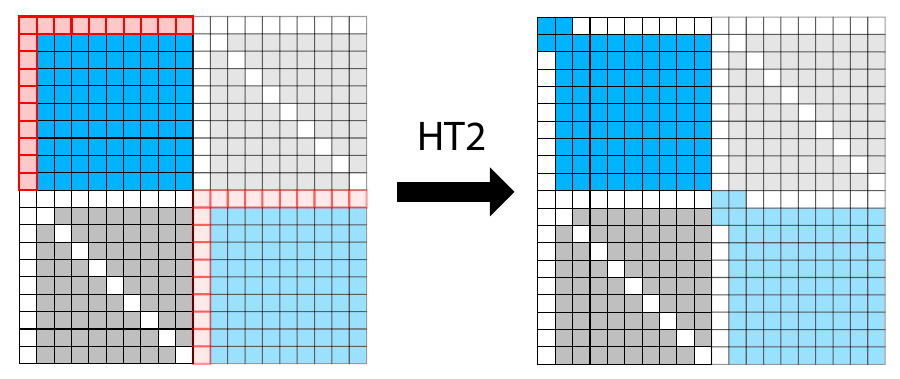}
\end{align}
The transformation reads
\begin{align}
\left( \begin{array}{cc} D & -E^\ast \\ E & D^\ast \end{array} \right) \leftarrow
\left( \begin{array}{cc} H_{D1} & 0 \\ 0 & H^\ast_{D1} \end{array}\right)\left( \begin{array}{cc} D & -E^\ast \\ E & D^\ast \end{array} \right) 
\left( \begin{array}{cc} H_{D1}^\dagger & 0 \\ 0 & H^T_{D1} \end{array}\right).
\label{ht2}
\end{align}
Repeating this procedure yields a block-diagonal matrix, whose diagonal blocks are Hermitian tridiagonal,
\begin{align}
\includegraphics[keepaspectratio, width=0.40\textwidth]{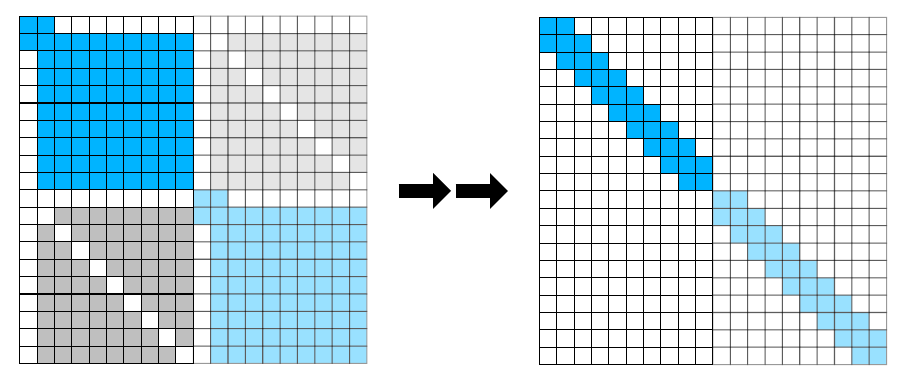}
\end{align}
The tridiagonal matrix can be efficiently diagonalized using the {\tt ZHBEV} subroutine in LAPACK.
In essence, the Paige--Van Loan algorithm takes advantage of the skew symmetry of the off-diagonal blocks and
clears them out via successive applications of the Householder transformations and Givens rotations.
It is observed that the operation count of this algorithm is roughly four times that of the diagonalization of $n\times n$ complex Hermitian matrices
(which is about a half of that of $2n\times 2n$ matrices).

The implementation of this algorithm is relatively straightforward.
For instance, the {\sc matlab} code that implements the above algorithm  has been presented by Loring \cite{Hastings2011AP,Loring2014NLAA}.
However, most of the computations are matrix--vector multiplications and outer-products of vectors, which are
computed by the Level 2 BLAS subroutines that have $O(n^2)$ floating point operations with $O(n^2)$ memory operations.
To attain the efficiency comparable to the state-of-the-art matrix diagonalization algorithms,
we ought to introduce a blocked algorithm such that most of the computations are mapped to the Level 3 BLAS subroutines
with $O(n^3)$ floating point operations and $O(n^2)$ memory operations.

\subsection{Kressner's compact WY-like representation}
Kressner's compact WY-like representation for structured matrices \cite{KressnerBook,Kressner2003BIT} is reviewed in this section.
In the state-of-the-art implementations of matrix diagonalization (e.g., {\tt ZHEEV}), the blocked algorithms are used
so that most of the computations can be done by the Level 3 BLAS subroutines \cite{Dongarra1989JCAM,Joffrain2006ACM}.
These modern algorithms are based on the fact that the product of Householder matrices can be compactly represented by the
so-called compact WY representation \cite{Bischof1987SIAM,Schreiber1989SIAM}:
\begin{align}
&Q^\dagger_k = H^\dagger_1 H^\dagger_2 \cdots H^\dagger_k  = 1 + W_kT_kW_k^\dagger, 
\end{align}
where $W\in \mathbb{C}^{n\times k}$ and $T\in \mathbb{C}^{k\times k}$ are recursively defined as
\begin{subequations}
\begin{align}
&T_{k} = \left(
\begin{array}{cc}
T_{k-1} & -\beta^\ast_k T_{k-1} W_{k-1}^\dagger v_k \\
0 & -\beta^\ast_k
\end{array}
\right),
\\
&W_{k} = \left(\begin{array}{cc} W_{k-1}& v_k\end{array}\right).
\end{align}
\end{subequations}
The accumulated transformation matrices for a panel of $A$ are used to
transform the rest of the matrix (see Sec.~\ref{algosec}), leveraging the efficiency of the Level 3 BLAS subroutines. 

Kressner has generalized the compact WY representation to the structured (Hamiltonian and skew-Hamiltonian) eigenvalue problems \cite{KressnerBook,Kressner2003BIT}.
The product of the transformation matrices are formally written as
\begin{align}
Q^\dagger_k = \bar{H}^T_{E1} G_1^\dagger \bar{H}^\dagger_{D1} \bar{H}^T_{E2} G_2^\dagger \bar{H}^\dagger_{D2}\cdots
\bar{H}^T_{Ek} G_k^\dagger \bar{H}^\dagger_{Dk}
\end{align}
where $G_k$ is defined in Eqs.~\eqref{givens1} and \eqref{givens2}, and $\bar{H}_{Xk}$ is
\begin{align}
\bar{H}_{Xk} = \left(\begin{array}{cc} H_{Xk} & 0 \\ 0 & H^\ast_{Xk} \end{array} \right).
\end{align}
with $X=D$ and $E$.
Note the complex conjugation of $\bar{H}_{E1}$ in Eq.~\eqref{ht1}. 
It has been shown in Ref.~\onlinecite{Kressner2003BIT} that $Q_k$ can be represented in a compact form,
\begin{align}
Q_k^\dagger =  \left(\begin{array}{cc} 1 + W_kT_kW_k^\dagger & W_k R_k S_k^\ast W_k^T \\
- W^\ast_k R_k^\ast S_k W_k^\dagger & 1 + W^\ast_k T^\ast_k W^T_k \end{array} \right),
\end{align}
in which the auxiliary matrices have the following structure:
\begin{subequations}
\begin{align}
&R_k = \left(\begin{array}{c} R^1 \\\hline R^2 \\\hline R^3 \end{array} \right),\\
&S_k = \left(\begin{array}{c|c|c} S^1& S^2& S^3\end{array} \right),\\
&T_k = \left(\begin{array}{c|c|c} T^{11} & T^{12} & T^{13} \\\hline
T^{21} & T^{22} & T^{23} \\\hline
T^{31} & T^{32} & T^{33}
\end{array} \right),\\
&W_k = \left(\begin{array}{c|c|c} W^1& W^2& W^3\end{array} \right),
\end{align}
\end{subequations}
with $W^1, W^2, W^3   \in \mathbb{C}^{n\times k}$.
All other matrices are in $\mathbb{C}^{k\times k}$ and upper-triangular.

The WY-like representation can be proven by induction with respect to $k$ \cite{Kressner2003BIT},
which in turn provides the way of constructing the auxiliary matrices.
The following equations are equivalent to those presented in Ref.~\onlinecite{Kressner2003BIT} except for complex conjugation and minor corrections.
For the first Householder transformation [Eq.~\eqref{ht1}], the updates are
\begin{subequations}
\label{step1}
\begin{align}
&R \leftarrow \left(\begin{array}{c} R^1 \\ 0 \\\hline R^2 \\\hline R^3 \end{array}\right),\\
&S \leftarrow \left(\begin{array}{cc|c|c} S^1 & -\beta_{Ek} SW^\dagger v^\ast_{Ek} & S^{2} & S^{3} \end{array}\right),\\
&T \leftarrow \left(
\begin{array}{cc|c|c}
T^{11} & -\beta_{Ek} T^{1:}W^\dagger v^\ast_{Ek} & T^{12} & T^{13} \\ 
0      & -\beta_{Ek}                             & 0      & 0      \\\hline
T^{21} & -\beta_{Ek} T^{2:} W^\dagger v^\ast_{Ek}& T^{22} & T^{23} \\\hline
T^{31} & -\beta_{Ek} T^{3:} W^\dagger v^\ast_{Ek}& T^{32} & T^{33} \\
\end{array}\right),\\
&W \leftarrow \left(\begin{array}{cc|c|c} W^1 & v^\ast_{Ek} & W^{2} & W^{3} \end{array}\right),
\end{align}
\end{subequations}
where we used a short-hand notation $T^{i:}$ for a row of blocks of $T$, i.e., $T^{i:} = (T^{i1}\,T^{i2}\,T^{i3})$.
Next, we update these matrices for the Givens rotation [Eq.~\eqref{givens}]:
\begin{subequations}
\label{step2}
\begin{align}
&R \leftarrow \left(\begin{array}{cc}
R^1 & T^{1:} W^\dagger e_{k+1} \\\hline
R^2 & T^{2:} W^\dagger e_{k+1} \\ 0 & 1 \\\hline
R^3 & T^{3:} W^\dagger e_{k+1}
\end{array}\right),\\
&S \leftarrow \left(\begin{array}{c|cc|c}
S^1 & S^{2} & \bar{c}_kS W^\dagger e_{k+1} & S^{3} \\
0   & 0     & -\bar{s}_k             & 0  
\end{array}\right),\\
&T \leftarrow \left(
\begin{array}{c|cc|c}
T^{11} & T^{12} & [\bar{s}_kR^1S^\ast W^T + \bar{c}_kT^{1:}W^\dagger] e_{k+1} & T^{13} \\\hline
T^{21} & T^{22} & [\bar{s}_kR^2S^\ast W^T + \bar{c}_kT^{2:}W^\dagger] e_{k+1} & T^{23} \\
0      & 0      & \bar{c}_k                                    & 0      \\ \hline
T^{31} & T^{32} & [\bar{s}_kR^3S^\ast W^T + \bar{c}_kT^{3:}W^\dagger] e_{k+1} & T^{33} 
\end{array}\right),\\
&W \leftarrow \left(\begin{array}{c|cc|c} W^1 & W^{2} & e_{k+1} & W^{3}
\end{array}\right),
\end{align}
\end{subequations}
in which we introduced $\bar{c}_k = \cos \theta_k - 1$ and $\bar{s}_k = (\sin \theta_k)^\ast$ with $\theta_k$ being
the rotation angle [see Eq.~\eqref{givens}]. $e_{k+1}$ is the ($k+1$)-th column of the unit matrix. 
Subsequently, the second Householder transformation [Eq.~\eqref{ht2}] updates these matrices as 
\begin{subequations}
\label{step3}
\begin{align}
&R \leftarrow \left(\begin{array}{c} R^1 \\\hline R^2 \\\hline R^3 \\ 0 \end{array}\right),\\
&S \leftarrow \left(\begin{array}{c|c|cc} S^1 & S^{2} & S^{3} & -\beta_{Dk}^\ast SW^\dagger v_{Dk}
\end{array}\right),\\
&T \leftarrow \left(
\begin{array}{c|c|cc}
T^{11} & T^{12} & T^{13} & -\beta_{Dk}^\ast T^{1:}W^\dagger v_{Dk} \\\hline
T^{21} & T^{22} & T^{23} & -\beta_{Dk}^\ast T^{2:}W^\dagger v_{Dk} \\\hline
T^{31} & T^{32} & T^{33} & -\beta_{Dk}^\ast T^{3:}W^\dagger v_{Dk} \\
0      & 0      & 0      & -\beta_{Dk}^\ast
\end{array}\right),\\
&W \leftarrow \left(\begin{array}{c|c|cc} W^1 & W^{2} & W^{3} & v^{Dk} \end{array}\right).
\end{align}
\end{subequations}
The matrices can be easily updated in each step by inserting a column and/or a row to the matrices.
Those after the first iteration are
\begin{subequations}
\begin{align} 
&R = \left( \begin{array}{ccc} -\beta_{E1} & 1 & 0\end{array}\right)^T\\
&S = \left( \begin{array}{ccc} 0 & -\bar{s}_1 & \bar{s}_1\beta_{D1}^\ast \end{array}\right) \\
&T = \left( \begin{array}{ccc} -\beta_{E1} & -\bar{c}_1\beta_{E1} &  \beta_{E1}\beta_{D1}^\ast (v_{E1}^T v_{D1} + \bar{c}_1 ) \\
0 & \bar{c}_1 &   -\bar{c}_1  \beta_{D1}^\ast \\
0 & 0 & -\beta_{D1}^\ast \\ 
\end{array}\right),
\end{align}
\end{subequations}
in which it is assumed that the second element of the Householder vectors is set to 1 [Eq.~\eqref{house}]. 

\subsection{Blocked updates and miscellaneous optimization\label{algosec}}
In the standard algorithms for Hermitian diagonalization, one can show that
\begin{align}
A_k = Q_k A Q_k^\dagger = (1+ W_kT_kW_k^\dagger)^\dagger (A + Y_kW_k^\dagger), 
\label{update}
\end{align} 
where $Y_k = AW_kT_k$.
Efficient programs accumulate $Q$ for a block of columns, or a panel, to update the entire matrix $A$ with Eq.~\eqref{update}
at once. The details are found in Ref.~\onlinecite{KressnerBook}.

We use a similar formula,
\begin{align}
\left(\begin{array}{c} D_k \\ E_k \end{array}\right)
&= \left(\begin{array}{cc} 1 + W_kT_kW_k^\dagger & W_k R_k S_k^\ast W_k^T \\
- W^\ast_k R_k^\ast S_k W_k^\dagger & 1 + W^\ast_k T^\ast_k W^T_k \end{array} \right)^\dagger\nonumber\\
&\times \left(\begin{array}{c} D+ (Y^D_k + Z^E_k)W^\dagger  \\ E + (Y^E_k - Z^D_k)W^\dagger  
\end{array}\right),
\label{panelupdate}
\end{align}
where $Y^D_k$ and $Z^D_k$ are in $\mathbb{C}^{n\times 3k}$ and defined as 
\begin{subequations}
\begin{align}
&Y_k^D = DW_kT_k, \\
&Z_k^D = D^\ast W^\ast_kR^\ast_k S_k.
\end{align}
\end{subequations}
$Y_k^E$ and $Z_k^E$ are likewise defined with $E$.
$Y$ and $Z$'s are computed in a three-step procedure as follows. After the first Householder transformation, they are
\begin{subequations}
\label{step1y}
\begin{align}
&Y^D \leftarrow \left(\begin{array}{cc|c|c} Y^{1D} & Y^D x -\beta_{Ek} Dv_{Ek}^\ast & Y^{2D} & Y^{3D} \end{array}\right),\\
&Z^D \leftarrow \left(\begin{array}{cc|c|c} Z^{1D} & Z^D x & Z^{2D} & Z^{3D} \end{array}\right),\\
& x = -\beta_{Ek} W^\dagger v_{Ek}^\ast.
\end{align}
\end{subequations}
Note that the auxiliary vector $x$ is computed in the update of $T$ and $S$ in Eq.~\eqref{ht1}.
They are transformed by the Givens rotation as
\begin{subequations}
\label{step2y}
\begin{align}
& Y^D \leftarrow \left(\begin{array}{c|cc|c} Y^{1D} & Y^{2D} & \bar{c}Y^D y + \bar{s}Z^{D\ast} y^\ast + \bar{c} D e_{k+1} & Y^{3D} \end{array}\right),\\
& Z^D \leftarrow \left(\begin{array}{c|cc|c} Z^{1D} & Z^{2D} & -\bar{s} Y^{D\ast} y^\ast + \bar{c} Z^D y - \bar{s} D^\ast e_{k+1} & Z^{3D} \end{array}\right),\\
& y = W^\dagger e_{k+1}.\label{yvec}
\end{align}
\end{subequations}
Finally, after the second Householder transformation, they become
\begin{subequations}
\label{step3y}
\begin{align}
&Y^D \leftarrow \left(\begin{array}{c|c|cc} Y^{1D} & Y^{2D} & Y^{3D} & Y^D z -\beta^\ast_{Dk} D v_{Dk} \end{array}\right),\\
&Z^D \leftarrow \left(\begin{array}{c|c|cc} Z^{1D} & Z^{2D} & Z^{3D} & Z^D z \end{array}\right),\\
&z = -\beta^\ast_{Dk} W^\dagger v_{Dk}.
\end{align}
\end{subequations}
$Y^E$ and $Z^E$ are computed similarly.

\begin{figure*}[t]
\includegraphics[keepaspectratio,width=0.9\textwidth]{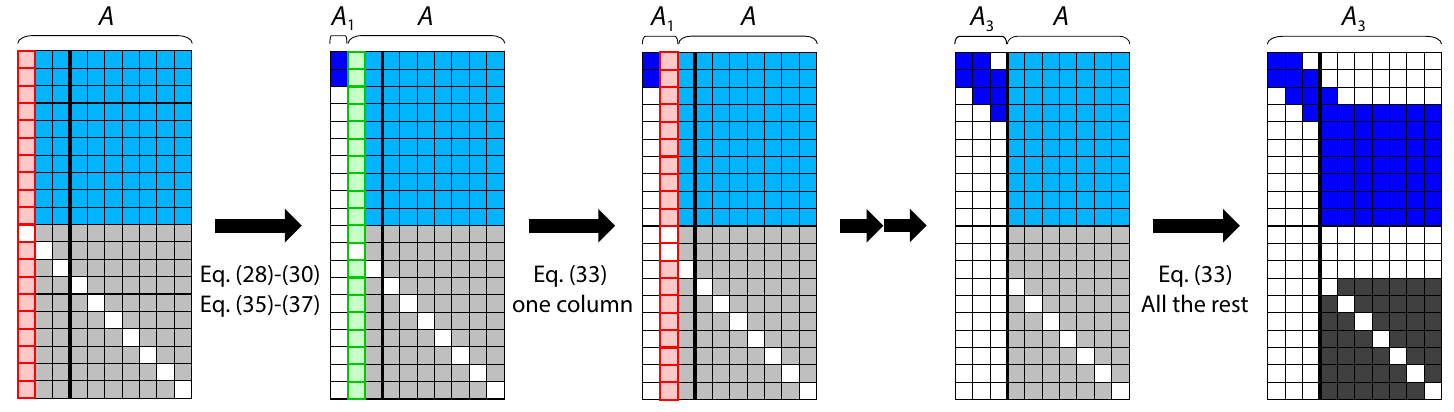}
\caption{Schematic representation of the blocked Paige--Van Loan algorithm ($n_b =3$). The last step is a dominant step, which is computed by the Level 3 BLAS subroutines.
This procedure is repeated until it reaches the last column and row.\label{kressner}}
\end{figure*}

The block update algorithm is sketched in Figure~\ref{kressner}.
First, we construct $R^1$, $S^1$, $T^1$, and $W^1$ using Eqs.~\eqref{step1}--\eqref{step3} while accumulating
$Y^{1D}$, $Y^{1E}$, $Z^{1D}$, and $Z^{1E}$ using Eqs.~\eqref{step1y}--\eqref{step3y}.
The first column is transformed to that in a tridiagonal form during this process.
Next, Eq.~\eqref{panelupdate} is applied to the second column, transforming it to the corresponding column of $A_1 = Q_1 AQ^\dagger_1$.
We then update all the auxiliary matrices and transform the second column to a tridiagonal form.
This procedure is repeated until we accumulate the matrices for $n_b$ columns. 
Finally, Eq.~\eqref{panelupdate} is used to update the rest of the matrix to yield $A_{n_b}$, which 
is efficiently performed by the Level 3 BLAS subroutines.
The problem size is then reduced from $n$ to $n-n_b$.
The above steps are recursively executed to complete the tridiagonalization.

\begin{figure*}[t]
\includegraphics[keepaspectratio,width=0.85\textwidth]{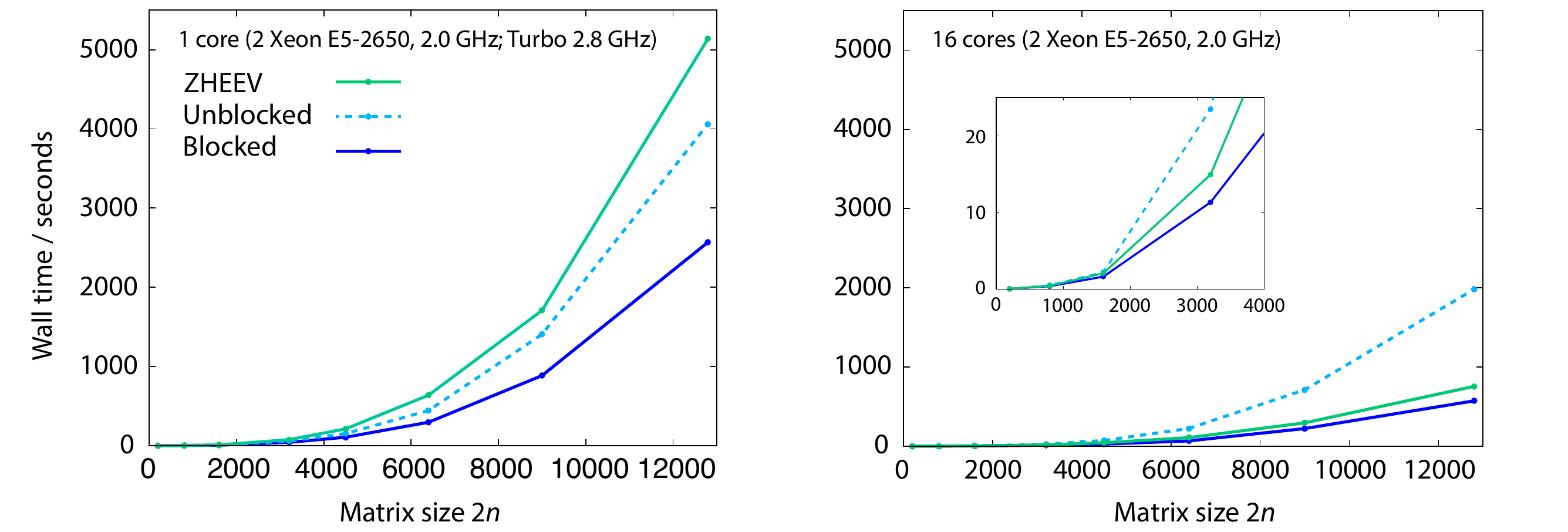}
\caption{Wall time for diagonalizing quaternionic matrices using the blocked and unblocked Paige--Van Loan algorithms, in comparison to MKL's {\tt ZHEEV} implementation
that ignores the quaternionic structure. 1 (left) and 16 (right) CPU cores were used, respectively. See text for details. \label{scaling}}
\end{figure*}

At this point, it is worth noting that several optimizations are possible in the implementation of above formulas. 
First, all of the elements in the first row of $W$ is zero, and therefore, the first row can be removed from the computation.
The first rows of $Y$'s and $Z$'s can also be omitted.
$W^2$ is then the first $n_b$ columns of the unit matrix,
which can be computed easily on the fly without storing it. 
Second, because all of the elements of $W^{2\dagger} e_{k+1}$ in Eq.~\eqref{yvec} are identically zero,
we do not need to keep $Y^2$ and $Z^2$'s except for the latest column; this column is the only one that
contributes to the panel update Eq.~\eqref{panelupdate} for the untransformed part of the matrix.
Therefore we overwrite $Y^2$ and $Z^2$'s in every iteration to reduce the memory requirement.
All of the elements of $W^{2\dagger} v_{Ek}^\ast$ in Eq.~\eqref{ht1} are zero, whereas the contributions from $W^{2\dagger} v_{Dk}$ in Eq.~\eqref{ht2} are trivial to compute. 
Finally, $W^1$ and $W^3$ (and other quantities) are stored in contiguous memory so that matrix--matrix multiplications can be fused. 
The memory size required for the auxiliary matrices after these optimizations is found to be around $11n_b n$ (note that
the optimal cache size for {\tt ZHEEV} is roughly $2n_b n$).

Our implementation takes advantage of efficient functions in BLAS and LAPACK as much as possible.
The Householder transformation is generated by {\tt ZLARFG}, while the Givens rotation parameters are calculated by {\tt ZLARTG}.
At the end of the algorithm, the $n\times n$ tridiagonal matrix is diagonalized by {\tt ZHBEV}.
Matrix--vector and matrix--matrix multiplications are performed by {\tt ZTRMV}, {\tt ZGEMV}, and {\tt ZGEMM} (or {\tt ZGEMM3M} when the Intel Math Kernel Library is used), respectively.
In the unblocked code, {\tt ZGERC} and {\tt ZGERU} are frequently used to calculate the outer product of vectors.
The Level 2 and 3 BLAS subroutines are responsible for shared-memory parallelization in our program.
The program was written in C++.

To our knowledge, this is the first implementation of the blocked Paige--Van Loan algorithm for quaternionic matrices,
whereas that for general real (skew-)Hamiltonian matrices has been implemented in the HAPACK package \cite{Benner2006ACM}.
Though diagonalization of a $2n\times 2n$ quaternionic matrix can be mapped to diagonalization of a $4n\times 4n$ skew-Hamiltonian matrix \cite{Benner1999ETNA},
direct solution in complex arithmetic is more efficient because doubling the size increases the cost of diagonalization by a factor of 8,
while diagonalization in complex arithmetic is only 3--4 times more expensive than that in real arithmetic (note that matrix multiplication
in complex arithmetic is 3 or 4 times more expensive than that in real arithmetic using {\tt ZGEMM3M} or {\tt ZGEMM} routines). 

\section{Timing results}
All of the timing data below were obtained on a computer node equipped with two Xeon E5-2650 2.0~GHz with the maximum Turbo Boost frequency 2.8~GHz
(Sandy Bridge, 8 cores each, 16 cores in total).
Intel Math Kernel Library (version 11.3, 2016.0.109) was used for BLAS and LAPACK functions with threading support.
The compiler used was g++ 5.2.0 with the optimization flags ``-O3 -DNDEBUG -mavx."
Figure~\ref{scaling} summarizes the timing for diagonalizing random matrices with the quaternionic structure using the blocked and unblocked algorithms described above ($n_b = 20$ was used).
The timing is compared with MKL's {\tt ZHEEV}, which does not preserve the structure of the eigenvectors. 
It is apparent that the use of the blocked algorithm is essential to achieving high efficiency comparable to the state-of-the-art implementation of standard diagonalization.
Diagonalizing a $12800\times 12800$ matrix took 9.5, 33.0, and 12.6 minutes using the blocked and unblocked versions of our code and MKL's {\tt ZHEEV} using 16 CPU cores.
When only 1 CPU core is used, the wall times were 42.8, 67.7, and 85.6 minutes, respectively.
Note that the calculations with 1 CPU core were accelerated by the CPU's Turbo Boost (up to 2.8~GHz).

The shared memory parallelization of our code appears less effective than that in the MKL implementation of {\tt ZHEEV},
indicating that there is still some room for improvement.
This is in part because our program is only threaded by the underlying BLAS functions and because the above blocked algorithm involves considerably more matrix--vector multiplications
in the computation of the WY-like representation than the one used in the standard diagonalization.
Although our new implementation does outperform MKL's {\tt ZHEEV} with 16 CPU cores by 25--30\%,
the detailed comparison with {\tt ZHEEV} is not straightforward because of the difference in design of the two programs.

\begin{figure}
\includegraphics[keepaspectratio,width=0.42\textwidth]{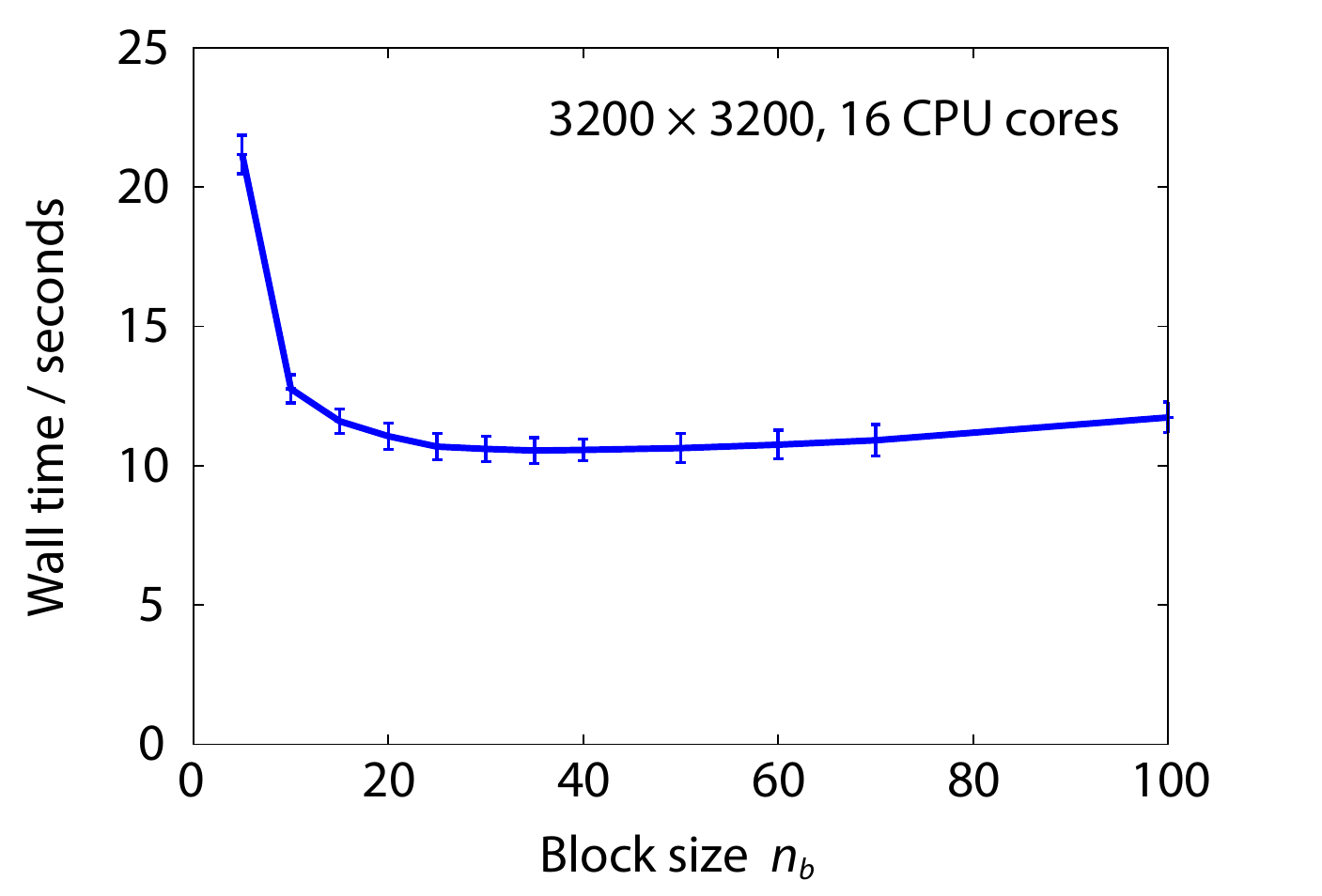}
\caption{Wall time as a function of the block size for diagonalizing random $3200\times 3200$ matrices with the quaternionic structure.
The error bars are derived from 50 repeated calculations.\label{tilefig}}
\end{figure}

The dependence of the efficiency with respect to the block size $n_b$ is presented in Figure~\ref{tilefig} for $3200\times 3200$ matrices
using 16 CPU cores. The timing improves dramatically till $n_b=20$, has the minimum 
at around $n_b= 35$, and then deteriorates slowly as $n_b$ becomes larger.
The initial improvement with respect to $n_b$ is attributed to the increased efficiency of the Level 3 BLAS subroutines with the increased panel sizes.
The deterioration with large $n_b$ is due to the use of the Level 2 BLAS subroutines within each panel [Eqs.~\eqref{step1}--\eqref{step3} and \eqref{step1y}--\eqref{step3y}],
although it competes with further efficiency improvement of the Level 3 BLAS subroutines in Eq.~\eqref{panelupdate}.
The optimal $n_b$ is expected to be machine dependent (e.g., L2 and L3 cache sizes).

\section{Conclusions}
We have implemented a blocked version of the Paige--Van Loan algorithm into an efficient program,
following the earlier work by Kressner \cite{Kressner2003BIT}.
Shared-memory parallelization is performed by the underlying BLAS and LAPACK library.
When a single CPU core is used, the elapsed time for structure-preserving diagonalization using this program is about a half of that using {\tt ZHEEV} without symmetry treatment;
with 16 CPU cores, the program still outperforms the MKL's {\tt ZHEEV} implementation by about 25--30\%. 
The source code is publicly available under the FreeBSD license \cite{zquatev} so that it can be integrated into any large-scale relativistic quantum chemistry program packages. 
The program has been interfaced to the {\sc bagel} package \cite{bagel}.
The parallel implementation will be investigated in the near future. 

\begin{acknowledgments}
This article is part of the special issue in honor of the 60th birthday of Hans J{\o}rgen Aa.~Jensen.
The author thanks Jack Poulson for helpful suggestions and Terry Loring for useful comments. 
This work has been supported by the NSF CAREER Award (CHE-1351598). 
The author is an Alfred P. Sloan Research Fellow.
\end{acknowledgments}

\end{document}